\documentclass[twocolumn]{aastex62}

\usepackage{amsmath}
\usepackage{cases}
\usepackage[shortlabels]{enumitem}

\newcommand{\twopartdef}[4]
{
	\left\{
		\begin{array}{ll}
			#1 & \mbox{if } #2 \\ \\
			#3 & \mbox{if } #4
		\end{array}
	\right.
}

\received{December 22, 2018}
\revised{February 26, 2019}
\accepted{February 27, 2019}
\submitjournal{ApJ Letters}

\shorttitle{Exomoon hide and seek}
\shortauthors{Martin, Fabrycky \& Montet}
\begin{document}

\title{Transits of Inclined Exomoons --- Hide and Seek and an Application to Kepler-1625}

\correspondingauthor{David V. Martin}
\email{davidmartin@uchicago.edu}

\author[0000-0002-7595-6360]{David V. Martin}
\affil{Department of Astronomy and Astrophysics, University of Chicago, 5640 S Ellis Ave, Chicago, IL 60637, USA}
\affil{Fellow of the Swiss National Science Foundation}

\author[0000-0003-3750-0183]{Daniel C. Fabrycky}
\affil{Department of Astronomy and Astrophysics, University of Chicago, 5640 S Ellis Ave, Chicago, IL 60637, USA}

\author[0000-0001-7516-8308]{Benjamin T. Montet}
\affil{Department of Astronomy and Astrophysics, University of Chicago, 5640 S Ellis Ave, Chicago, IL 60637, USA}
\affil{NASA Sagan Fellow}

\begin{abstract}

A Neptune-sized exomoon candidate was recently announced by \citet{Teachey:2018kj}, orbiting a 287-day gas giant in the Kepler-1625 system. However, the system is poorly characterized and needs more observations to be confirmed, with the next potential transit in May 2019. In this letter, we aid observational follow up by analyzing the transit signature of exomoons. We derive a simple analytic equation for the transit probability and use it to demonstrate how exomoons may frequently avoid transit if their orbit is larger than the stellar radius and sufficiently misaligned. The nominal orbit for the moon in Kepler-1625 has both of these characteristics, and we calculate that it may only transit $\approx40\%$ of the time. This means that $\approx$six non-transits would be required to rule out the moon's existence at $95\%$ confidence. When an exomoon's impact parameter is displaced off the star, the planet's impact parameter is displaced the other way, so larger planet transit durations are typically positively correlated with missed exomoon transits. On the other hand, strong correlations do not exist between missed exomoon transits and transit timing variations of the planet. We also show that nodal precession does not change an exomoon's transit probability and that it can break a prograde-retrograde degeneracy.

\end{abstract}


\section{Introduction}
\label{sec:intro}

In the solar system our understanding of the planets is enriched by our understanding of their moons. The Moon is thought to influence Earth's habitability \citep{Laskar:1993fg}. The Galilean moons help constrain the early evolution of Jupiter \citep{Heller:2015yu,Ronnet:2018nm}. The equatorial alignment of Uranus's moons helps us understand the origin of the planet's tilt \citep{Kegerreis:2018fe}. As a community we would benefit immensely from conducting similar science for moons of extrasolar planets (exomoons).

%, as the large diversity of diskoveries has challenged how we think planets form and evolve. 

Detecting analogs of the solar system moons is challenging due to their small size. Photometry is thought to be the most promising technique \citep{Kipping:2009et}, either through observing individual moon transits \citep{Sartoretti:1999tr}, multiple averaged moon transits \citep{Simon:2012lr,Heller:2014ro,Teachey:2017hf} or inferring the moon's existence based on the planet's transit timing variations (TTVs) and transit duration variations (TDVs) \citep{Sartoretti:1999tr,Kipping:2009aa,Kipping:2009bb,Kipping:2011tr,Heller:2016tr}. Other techniques with potential include gravitational microlensing \citep{Bennett:2014jq,Hwang:2018zp} and observations of self-luminous giant exoplanets to detect a variation in polarization \citep{Sengupta:2016uf} or in radial velocity \citep{Vanderburg:2018ur}.

The most plausible exomoon to date is in the Kepler-1625 system. The planet (Kepler-1625b) itself is unremarkable: a gas giant on a 287-day orbit. The surprise, however, is the size of the moon (Kepler-1625b-i, as it is potentially similar in mass and radius to Neptune. Such a large moon is without precedent in our solar system, but one must remember that so were the first exoplanet diskoveries. 

\begin{table*}
\caption{Parameters of the Kepler-1625 Exomoon candidate system}\label{tab:params}
\vspace{0.5cm}
\begin{minipage}{1\textwidth}
\begin{minipage}{1\textwidth}
\end{minipage}
\begin{minipage}{0.52\textwidth}
%\begin{table}[]\label{tab:params}
%\caption{Parameters of the Kepler-1625 Exomoon candidate system}
\begin{tabular}{llllll}
\hline
Param.   & Unit            & Value & $1\sigma$ Min & $1\sigma$  Max & Note \\ 
\hline\hline
\multicolumn{6}{c}{Host star}  \\
\hline
$m_{\star}$ & $(M_{\odot})$   & 1.04      & 0.98    & 1.12    & \\
$R_{\star}$ & $(R_{\odot})$   & 1.73      & 1.51    & 1.97    & \\
\hline
\multicolumn{6}{c}{Planet}  \\
\hline
$m_{\rm P}$ & $(M_{\rm Jup})$ & 6.85      & 1.2    & 12.5    & (a) \\
$R_{\rm P}$ & $(R_{\rm Jup})$ & 1.04      & 0.90    & 1.18    & \\
$T_{\rm P}$ & (days)          & 287.37278      & 287.37213    & 287.37353    & \\
$a_{\rm P}$ & (au)            & 0.87      &  0.85   & 0.89   & (b) \\
$b_{\rm P}$ &            & 0.104      & 0.038    & 0.188   & \\
$I_{\rm P}$ & (deg)            & 89.94      & 89.88    & 89.98   & (c)\\
$\Omega_{\rm P}$ & (deg)            & 0      & 0    & 0   & (c)\\
\hline
\multicolumn{6}{c}{Moon}  \\
\hline
$m_{\rm M}$ & $(M_{\oplus})$   & 36.2      & 4.4    & 68    & (a)\\
$R_{\rm M}$ & $(R_{\oplus})$  & 4.90      & 4.18    & 5.69    & \\
$T_{\rm M}$ & (days)          & 22      & 13    & 39    & \\
$a_{\rm M}$ & (au)            & 0.022      & 0.017    & 0.030   & (d)\\
$I_{\rm M}$ & (deg)            & 42      & 24    & 57   & (e)\\
$\Omega_{\rm M}$ & (deg)            & 0      & -83    & 142   & (e)\\
\hline
\multicolumn{6}{c}{Relative orbit}  \\
\hline
$|90-I_{\rm M}|$ & (deg)            & 48      & 33    & 66   & (f)\\
\end{tabular}
%\end{table}
\end{minipage}
\begin{minipage}{0.45\textwidth}
{\bf Parameter key:} $m$: mass, $R$: radius, $T$: period, $a$: semi-major axis, $b$: impact parameter, $I$: inclination, $\Omega$: longitude of the ascending node. 
\vspace{0.1cm}
%{\bf Notes:}
\begin{enumerate}[(a)]
\item No nominal value is given for the planet or moon mass, only upper and lower bounds, so the value that we provide here is simply an average.
\item \citet{Teachey:2018kj} gave $a_{\rm P}=0.98^{+0.14}_{-0.13}$ au but this is inconsistent with their values for $T_{\rm P}=287$ days and $M_{\star}=0.98^{+0.08}_{-0.06}$ $M_{\odot}$ so we recalculate $a_{\rm P}$ and our value matches \citet{Heller:2018oe}.
\item $I_{\rm P}$ is not given by \citet{Teachey:2018kj}; calculated from our value of $a_{\rm P}$ and the given values of $b_{\rm P}$. $\Omega_{\rm P}=0^{\circ}$ arbitrarily because transits are not sensitive to both $\Omega_{\rm P}$ and $\Omega_{\rm M}$ individually, only  $\Delta \Omega$.
\item Not given by \citet{Teachey:2018kj}; calculated from their values of $a_{\rm M}/R_{\rm P}=45^{+10}_{-5}$.
\item We take $I_{\rm M}$ and $\Omega_{\rm M}$ to be calculated with respect to the observer, although we note that $\Omega_{\rm M}$ is essentially unconstrained by the data, with a $225^{\circ}$ $1\sigma$ confidence interval. The inclination value is also modulo $90^{\circ}$, i.e. a degeneracy exists.
\item Equivalent to $\Delta I$ from Eq.~\ref{eq:Delta_I} with $\Omega_{\rm M}=0^{\circ}$ and $I_{\rm P}=90^{\circ}$. We use this as the moon's mutual inclination because \citet{Teachey:2018kj} did not give a value and $\Omega_{\rm M}$ is so poorly constrained.
\end{enumerate}
\end{minipage}
\end{minipage}
\begin{minipage}{1\textwidth}
\vspace{0.3cm}
\makebox[\linewidth]{\rule{0.83\paperwidth}{0.4pt}}
\end{minipage}
\end{table*}

The moon was originally suspected based on three planet transits within the original {\it Kepler} mission \citep{Teachey:2017hf,Heller:2018oe}. Asymmetries in the transit profile teased the presence of a moon, but neither TTVs nor TDVs were detected to confirm it. The moon's existence became more likely after a fourth planetary transit was captured by the Hubble Space Telescope (HST) \citep{Teachey:2018kj}. The planet transit was 70 minutes early, although no TDV was detected. Furthermore, there is a shallow dip in the light curve after the egress of the planet transit: a potential moon transit. Table~\ref{tab:params} contains basic system parameters used in our letter, but we refer the reader to \citet{Teachey:2018kj} for significantly more detail.

\begin{figure*}
\gridline{\fig{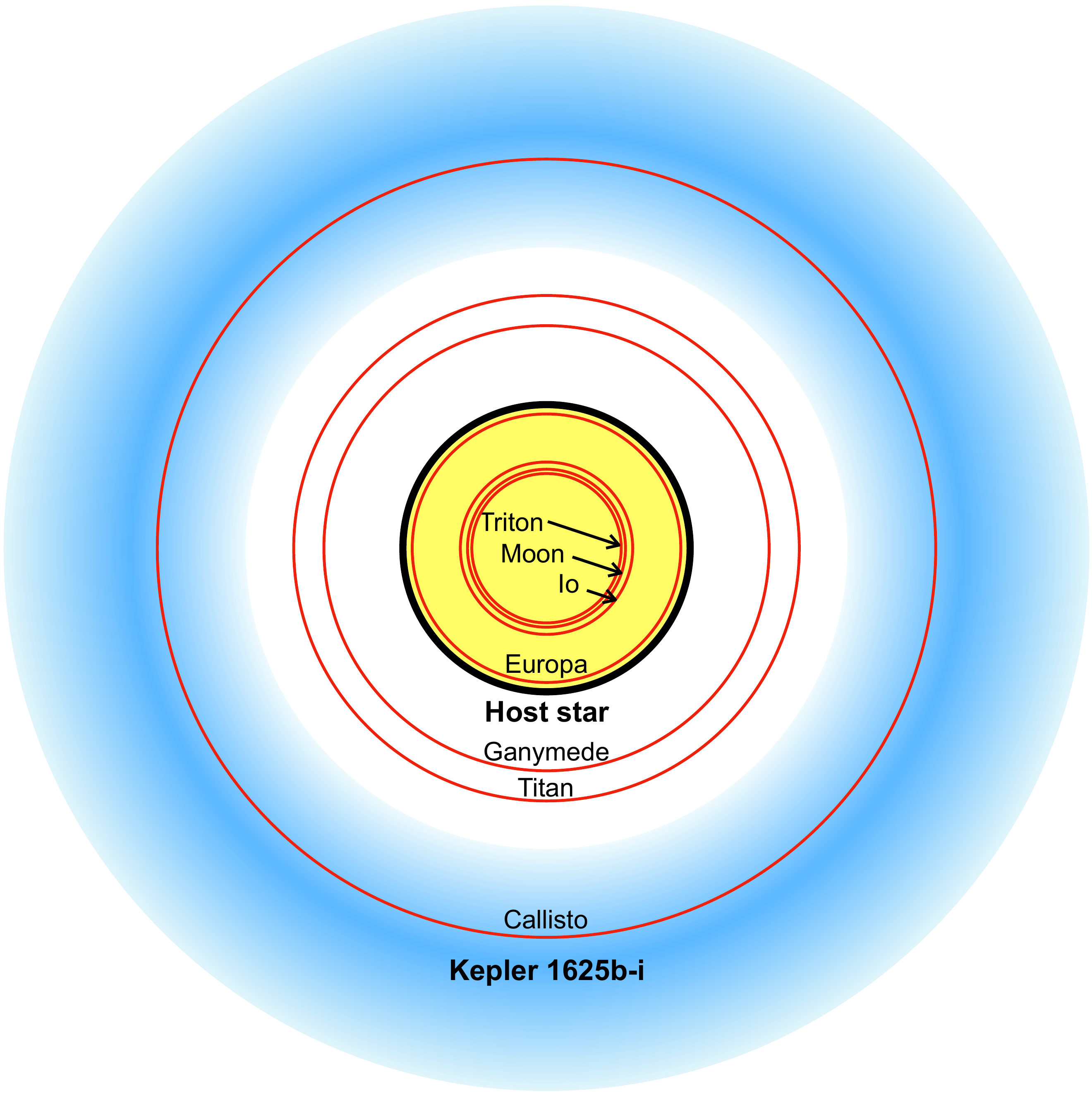}{0.48\textwidth}{}\label{fig:moon_orbit_sizes}
          \fig{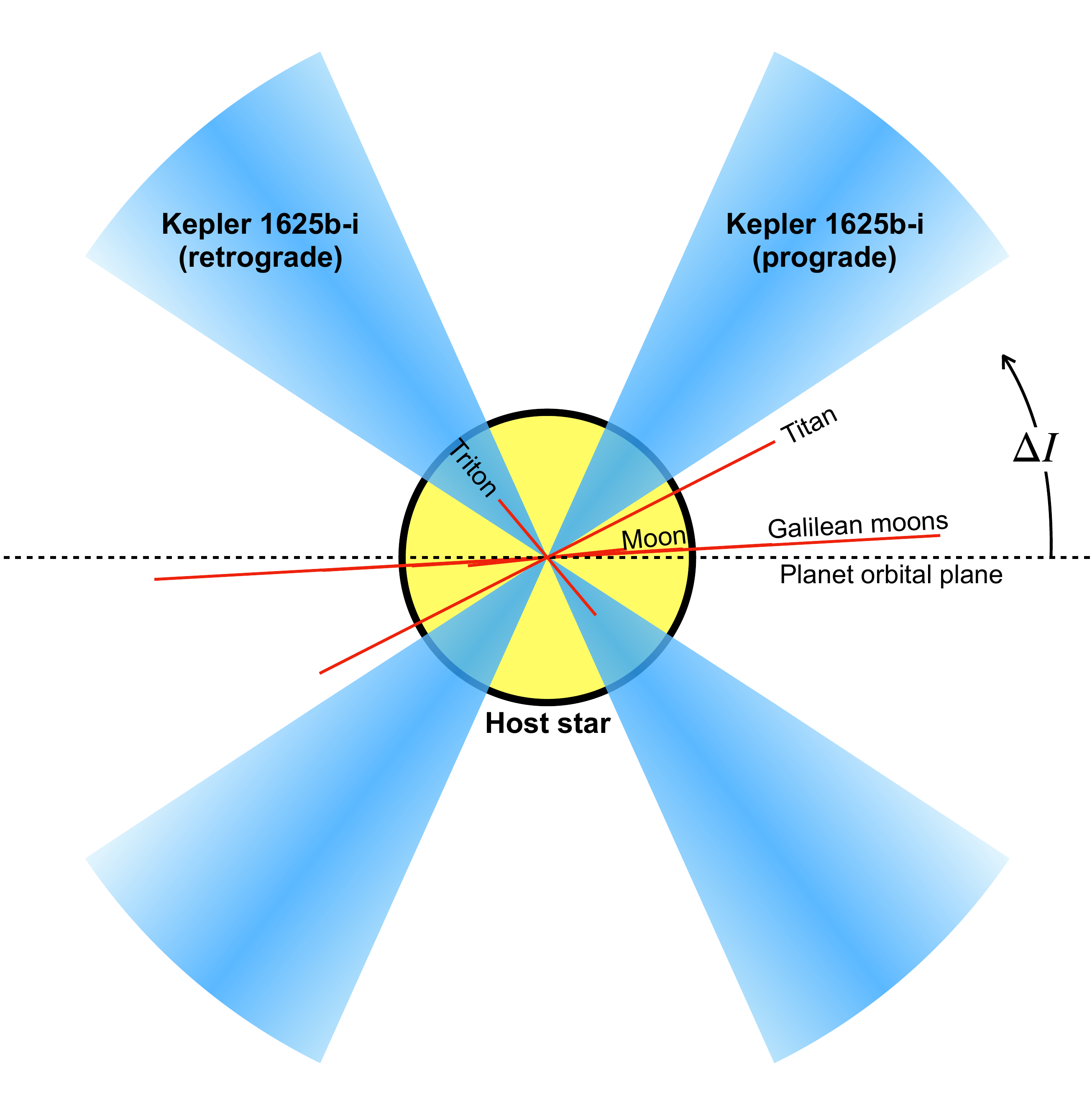}{0.49\textwidth}{}\label{fig:moon_orbit_inclinations}
          }
\caption{Left: orbits of the seven most massive solar system moons (red) and the exomoon candidate Kepler-1625b-i (the shaded blue region denotes the $1\sigma$ $a_{\rm M}$ error bars) compared with the host star disk, ignoring eccentricity. Right: mutual inclination ($\Delta I$) measured counter-clockwise from the planet orbital plane (black dashed horizontal line) to the moon orbital plane. The massive solar system moons are shown as individual red lines, although most closely overlap. For Kepler-1625 we estimate $\Delta I\approx90^{\circ}-I_{\rm M}$ from Eq.~\ref{eq:Delta_I} with $\Delta \Omega=0$ and $I_{\rm P}=90^{\circ}$. For the error in $\Delta I$ we take the given $1\sigma$ errors for $I_{\rm M}$. A blue shaded region shows the $1\sigma$ confidence interval and is mirrored for retrograde. Note that Titan is actually almost coplanar to its host Saturn's equator, but the planet is tilted by $\Delta I=27^{\circ}$ from its orbital plane. 
\label{fig:solar_system_moons}}
\end{figure*}

In this letter we are agnostic about the reality of this particular exomoon. Both \citet{Teachey:2018kj} and subsequent analysis by \citet{Heller:2019fg} encourage new observations in order to consider the moon confirmed. In this letter we aid such future observations by analyzing the detectability of exomoons, both in general and for Kepler-1625b-i specifically. We quantify previous intuition that some moons are not guaranteed to transit every time their host planet does \citep{Sartoretti:1999tr,Martin:2017oi}. Missed transits typically occur when the moon's orbit is both wider than the stellar diameter and significantly misaligned to the planet's orbital plane. The best-fitting, albeit loosely constrained orbit for Kepler-1625b-i has both of these characteristics. Furthermore, within our own solar system we know of Triton, which is on a highly misaligned, in fact retrograde, orbit (Fig.~\ref{fig:solar_system_moons}).

In this letter we derive an analytic transit probability for exomoons of transiting planets (Sect.~\ref{sec:setup}) which accounts for both misalignment and a dynamically varying exomoon orbit. We then test the correlation between the presence/absence of moon transits and the TTV and TDV signature of the planet (Sect.~\ref{sec:transit_timing}). We apply our work to both exomoons in general and the Kepler-1625 system specifically (Sect~\ref{sec:applications}). The letter ends with a brief discussion (Sect.~\ref{sec:discussion}).

\section{Exomoon transit probability}\label{sec:setup}

\subsection{Transit Geometry}\label{subsec:setup_transitgeometry}

The transit geometry is shown in Fig.~\ref{fig:basic_geometry}. The observer looks from the positive $z$-axis at the $(x,y)$ sky plane centered on the star. The planet orbit is modeled by a straight line from left to right (positive $x$ direction), vertically offset by the impact parameter $b_{\rm P}=a_{\rm P}\cos I_{\rm P}/R_{\star}$. This assumes $a_{\rm P} \gg R_{\star}$\footnote{Very tight-orbiting planets are thought unlikely to host moons anyway \citep{Namouni:2010tx}.} and $m_{\rm P} \gg m_{\rm M}$\footnote{Care must be taken when generalizing our work to  ``binary planets'' \citep{Lewis:2015mm}, although our work is likely applicable to ``moon-moons'' \citep{Forgan:2018do}, ``moon-moon-moons'' or indeed moon$^n$.} and throughout this letter we also assume circular orbits, i.e. $e_{\rm P}=e_{\rm M}=0$. The planet's orbit would be rotated clockwise by $\Omega_{\rm P}$, but we arbitrarily set $\Omega_{\rm P}=0$ as the transit geometry are only sensitive to $\Delta \Omega = \Omega_{\rm M} - \Omega_{\rm P}$.

The position of the moon at the time of the planet's transit midpoint across the star is fundamental to the transit phenomenom. Neglecting eccentricity, its projected orbit is an ellipse with major axis $a_{\rm M}$ and a minor axis $a_{\rm M}|\cos I_{\rm M}|$, rotated counter-clockwise by $\Omega_{\rm M}$  and offset vertically by $b_{\rm P}R_{\star}$:

\begin{figure*}  
\begin{center}  
\includegraphics[width=0.99\textwidth]{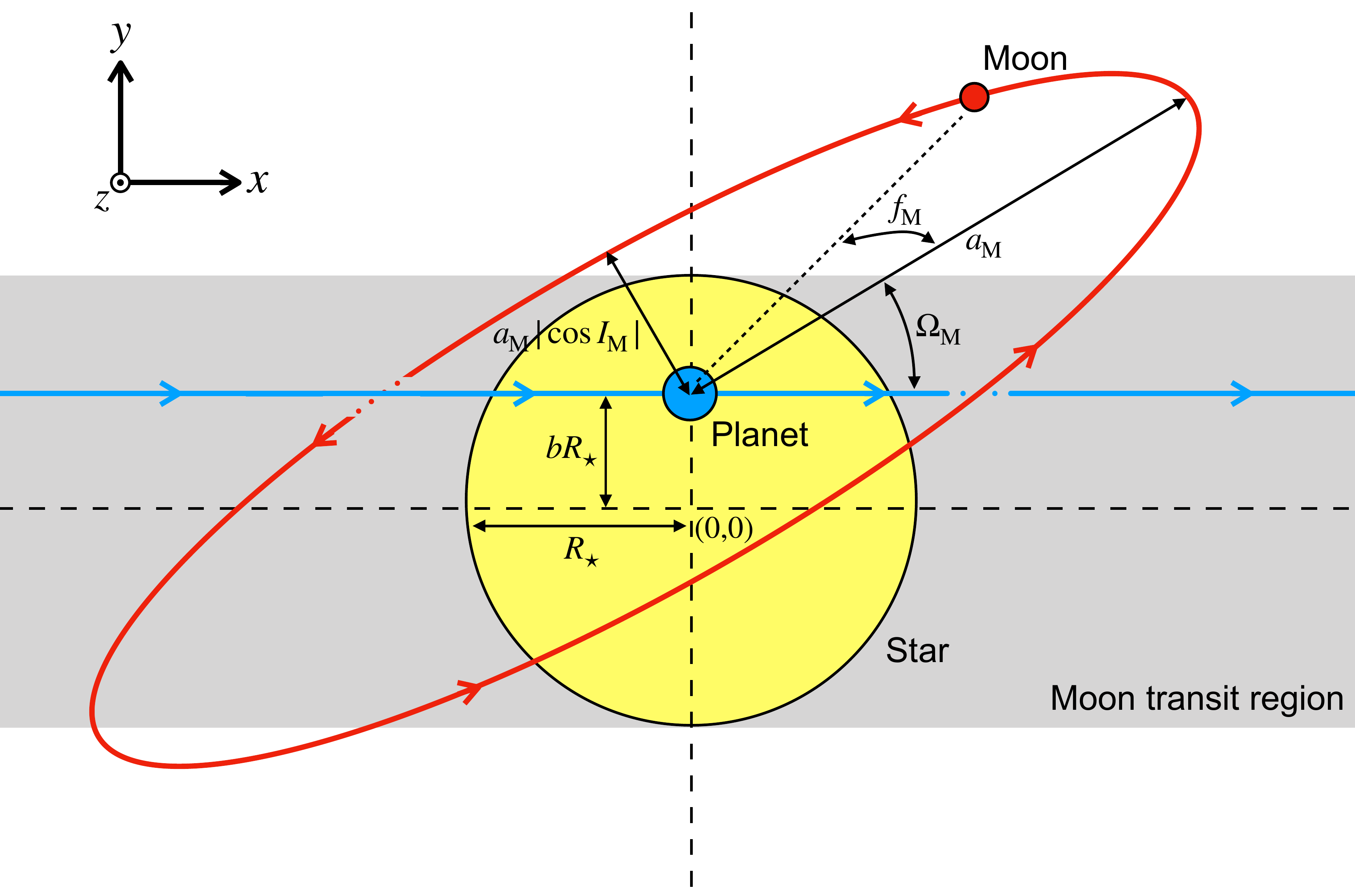}  
%\caption{Different type of planet orbits in binaries}
\caption{Observer's view of a transiting exoplanet (blue), its host star (yellow), and exomoon (red). Moons within the gray region will transit the star. Dotted regions of the moon and planet orbits show where those orbits pass behind the projected orbit of the other body.}
\label{fig:basic_geometry}
\end{center}  
\end{figure*}

\begin{equation}
\footnotesize
\label{eq:moon_orbit}
\left[\begin{array}{c}x_{\rm M}(f_{\rm M}) \\y_{\rm M}(f_{\rm M})\end{array}\right] = \left[\begin{array}{c} a_{\rm M}\left(\cos \Omega_{\rm M} \cos f_{\rm M} - \cos I_{\rm M} \sin \Omega_{\rm M} \sin f_{\rm M}\right)\\ a_{\rm M}\left(\sin \Omega_{\rm M} \cos f_{\rm M} + \cos I_{\rm M} \cos \Omega_{\rm M} \sin f_{\rm M}\right) + bR_{\star}   \end{array}\right],
\end{equation}

where $f_{\rm M}$ is the true anomaly of the moon. It is important to remember that $f_{\rm M}$ is the orbital phase of the moon defined within its orbital plane, {\it not} with respect to our $(x,y)$ coordinate system. In Fig.~\ref{fig:basic_geometry} the moon is misaligned and prograde with the planet's orbit and projects a counter-clockwise motion. In Fig.~\ref{fig:degeneracy} we however note that a degeneracy exists between prograde and retrograde moons (see Sect.~\ref{subsec:degeneracy}). The mutual inclination between the moon and the planet's orbit (not equator) is
\begin{equation}
\label{eq:Delta_I}
\cos \Delta I = \cos \Delta \Omega\sin I_{\rm M}\sin I_{\rm P} + \cos I_{\rm M} \cos I_{\rm P}.
\end{equation}

%\begin{figure*}
%\gridline{\fig{basic_geometry.pdf}{0.99\textwidth}{(a)}\label{fig:basic_geometry}}
%\gridline{\fig{degeneracy.pdf}{0.99\textwidth}{(b)}\label{fig:degeneracy}}
%\caption{(a) Observer's view of a transiting exoplanet (blue), its host star (yellow), and exomoon (red). Moons within the gray region will transit the star. Dotted regions of the moon and planet orbits show where those orbits pass behind the projected orbit of the other body. (b) An exoplanet orbit (blue) with two different exomoon orbits: the red solid line is prograde and coplanar to the planet, and the black dashed line is retrograde and misaligned to the planet. As seen by the observer (right) there is a degeneracy, as both moons has the same $(x,y)$ position and $(v_x,y_y)$ velocity, despite the side view (left) betraying a clear difference between the two orbits. $\mathbf{L}$ denotes various angular momentum vectors.\label{fig:geometry}}
%\end{figure*}

The moon will transit the star on a given planet transit when $ |y_{\rm M}(f_{\rm M})|<R_{\star}$. To make this criterion easier to solve, we consolidate the expression in Eq.~\ref{eq:moon_orbit} for $y_{\rm M}$ from two trigonometric functions of $f_{\rm M}$ to one:

\begin{equation} \label{eq:ym_new}
\begin{split}
y_{\rm M}(f_{\rm M}) & = a_{\rm M}\sqrt{\sin^2\Omega_{\rm M} + \cos^2I_{\rm M}\cos^2\Omega_{\rm M}}  \\
& \times \cos\left(f_{\rm M} - \arctan\left[{\frac{\cos I_{\rm M}}{\tan \Omega_{\rm M}}}\right] \right)+ bR_{\star} \\
 & \approx a_{\rm M}\left|\sin \Delta I\right|  \cos\left(f_{\rm M} - \arctan\left[{\frac{\cos I_{\rm M}}{\tan \Omega_{\rm M}}}\right] \right)+ bR_{\star},
\end{split}
\end{equation}
We note that whilst the second line of Eq.~\ref{eq:ym_new} contains $b_{\rm P}$, hence implying that $I_{\rm P}$ is not exactly $90^{\circ}$, the approximation  $\sqrt{\sin^2\Omega_{\rm M} + \cos^2I_{\rm M}\cos^2\Omega_{\rm M}}\approx |\sin \Delta I|$ is derived from Eq.~\ref{eq:Delta_I} using $I_{\rm P}= 90^{\circ}$. However, the end result is a negligible difference between the two lines in Eq.~\ref{eq:ym_new}.

The exomoon transit probability is calculated as the fraction of angles $f_{\rm M}$ that correspond to $|y_{\rm M}| < R_{\star}$. 
The phase shift of $\arctan[\cos I_{\rm M}/\tan\Omega_{\rm M}]$ in Eq.~\ref{eq:ym_new} does not affect this fraction, and hence we simplify Eq.~\ref{eq:ym_new} by defining $f'_{\rm M}=f_{\rm M}-\arctan[\cos I_{\rm M}/\tan\Omega_{\rm M}]$. The function $y_{\rm M} ( f'_{\rm M})$ is symmetric over $f'_{\rm M}=180^{\circ}$. Between 0 and $180^{\circ}$ we define the range of transits to be $[A,B]$, where

\begin{equation}
\footnotesize
\label{eq:A}
A = \twopartdef{0}{b_{\rm P}R_{\star} + a_{\rm M}|\sin\Delta I |< R_{\star}}{\arccos\left[\frac{R_{\star}(1-b_{\rm P})}{a_{\rm M}\sin\Delta I} \right]}{b_{\rm P}R_{\star} + a_{\rm M}|\sin\Delta I |> R_{\star}  \\  \hspace{2cm}}
\end{equation}
where the second condition occurs when the moon can miss transit {\it above} the star (with respect to the $y$-axis, and 
\begin{equation}
\footnotesize
\label{eq:B}
B = \twopartdef{180^{\circ}}{b_{\rm P}R_{\star} - a_{\rm M}|\sin\Delta I |> -R_{\star}}{\arccos\left[\frac{-R_{\star}(1+b_{\rm P})}{a_{\rm M}\sin\Delta I} \right]}{b_{\rm P}R_{\star} - a_{\rm M}|\sin\Delta I |< -R_{\star}}
\end{equation}
where the second condition occurs when the moon can miss transit {\it below} the star.

If $f'_{\rm M}$ (and hence $y_{\rm M}$) is static during the planet's transit then the exomoon transit probability is simply the ratio $p_{\rm M}=(B-A)/180^{\circ}$. However, this static assumption is only applicable when  $T_{\rm M}\gg\tau_{\rm P}$, where $\tau_{\rm P}$ is the planet's transit duration:
\begin{equation}
 \label{eq:tauP}
\tau_{\rm P} = \frac{T_{\rm P}}{\pi}\arcsin\left(\frac{R_{\star}\sqrt{1-b_{\rm P}^2}}{a_{\rm P}}\right),
\end{equation}
To approximately account for shorter-period moons we add to $p_{\rm M}$ the fraction of the orbit covered during the planet's transit: $\tau_{\rm P}/T_{\rm M}$. With this, our derived exomoon transit probability is

\begin{equation}
\label{eq:moon_transit_probability}
p_{\rm M} = \min\left[\frac{B-A}{180^{\circ}} + \frac{\tau_{\rm P}}{T_{\rm M}},1\right].
\end{equation}

\subsection{Orbital dynamics}\label{sec:setup_dynamics}

The orbit of an exomoon may be subject to various dynamical perturbations. When the moon and planet orbits are misaligned, one such effect is a nodal precession induced by the three-body interactions between the sun, planet, and moon. From \citet{Mardling:2010qw} the rate of precession is
\begin{align}
\label{eq:precession_period}
T_{\rm prec}= \frac{4}{3}\frac{m_{\rm P}+m_{\star}}{m_{\star}}\frac{T_{\rm P}^2}{T_{\rm M}}\frac{1}{\cos \Delta I}.
\end{align}

This effect may be quenched by a competing torque on the moon's orbit induced by the equatorial bulge of the planet. \citet{Burns:1986gd} calculate a critical moon semi-major axis, for which the dynamics of interior orbits are dominated by the planet's equatorial bulge:

\begin{equation}
a_{\rm M,crit} = \left(\frac{2J_2R_{\rm P}^2a_{\rm P}^3m_{\rm P}}{m_{\star}}\right)^{1/3},
\end{equation}
where $J_2$ is the first gravitational harmonic. See also  \citet{Boue:2006oi,Tremaine:2009tm} for more details. In this letter we are predominantly interested in moons that are long-period and misaligned (such that moon transits are sometimes missed) and planets that are short-period (so planet transits are more frequent). For such moons the dominant effect is a three-body nodal precession. The Earth's moon exhibits three-body nodal precession with a period of 17.9 years (according to Eq.~\ref{eq:precession_period}). For Kepler-1625 $a_{\rm M,crit}=0.008$ au, which is almost three times less than the nominal value $a_{\rm M}=0.022$ au, and hence we also expect three-body nodal precession in this system, with a calculated period of 20.5 years.

With respect to the orbital plane of the planet, which remains (essentially) fixed, nodal precession makes the moon orbit circulate at a constant rate given by Eq.~\ref{eq:precession_period}, whilst maintaining a constant mutual inclination $\Delta I$. With respect to the observer, \citet{Martin:2017oi} showed that $I_{\rm M}$ librates over time $t$ around the constant $I_{\rm P}$ according to

\begin{equation}
\label{eq:moon_incl_over_time}
I_{\rm M}(t) = \Delta I \cos\left[\frac{2\pi}{T_{\rm prec}}(t-t_0)\right] + I_{\rm P},
\end{equation}
where $t_0$ corresponds to $I_{\rm M,0}$. With respect to the observer, $\Omega_{\rm M}(t)$ also librates and can be calculated by combining Eqs.~\ref{eq:Delta_I} and \ref{eq:moon_incl_over_time}.

A complication to the nodal precession arises in highly misaligned orbits, such that $|90^{\circ} - \Delta I|\lesssim 50^{\circ}$. In such cases Kozai-Lidov cycles occur, which cause $\Delta I$ and $e_{\rm M}$ vary, even for initially circular orbits \citep{Lidov:1961ru,Lidov:1962kx,Kozai:1962qf}.

The expression for $y_{\rm M}$ in  Eq.~\ref{eq:ym_new}, {\it does} depend on the time-dependent quantities $I_{\rm M}$ and $\Omega_{\rm M}$. However, these quantities only phase shift $f_{\rm M}$  and {\it do not} change the fractional range of $f_{\rm M}$ corresponding to transits, which is why they could be ignored when calculating the quantities $A$ (Eq.~\ref{eq:A}) and $B$ (Eq.~\ref{eq:B}). These quantities are functions of $\Delta I$, but this is constant\footnote{To be precise, $\Delta I$ is only constant under the secular regime, i.e. when calculations are made that average over the orbital periods. There do exist short-term variations on the timescales of $T_{\rm M}$ and $T_{\rm P}$, but these are on order $\approx2\%$ variations.} for orbits that are circular and without  Kozai-Lidov cycles. Overall, we demonstrate that in our simplified setup the exomoon transit probability $p_{\rm M}$ is constant during the moon's precession period.

\begin{figure*}

\vspace{1cm}
\begin{minipage}{0.49\textwidth}
\begin{minipage}{0.9\textwidth}
\includegraphics[width=0.99\textwidth]{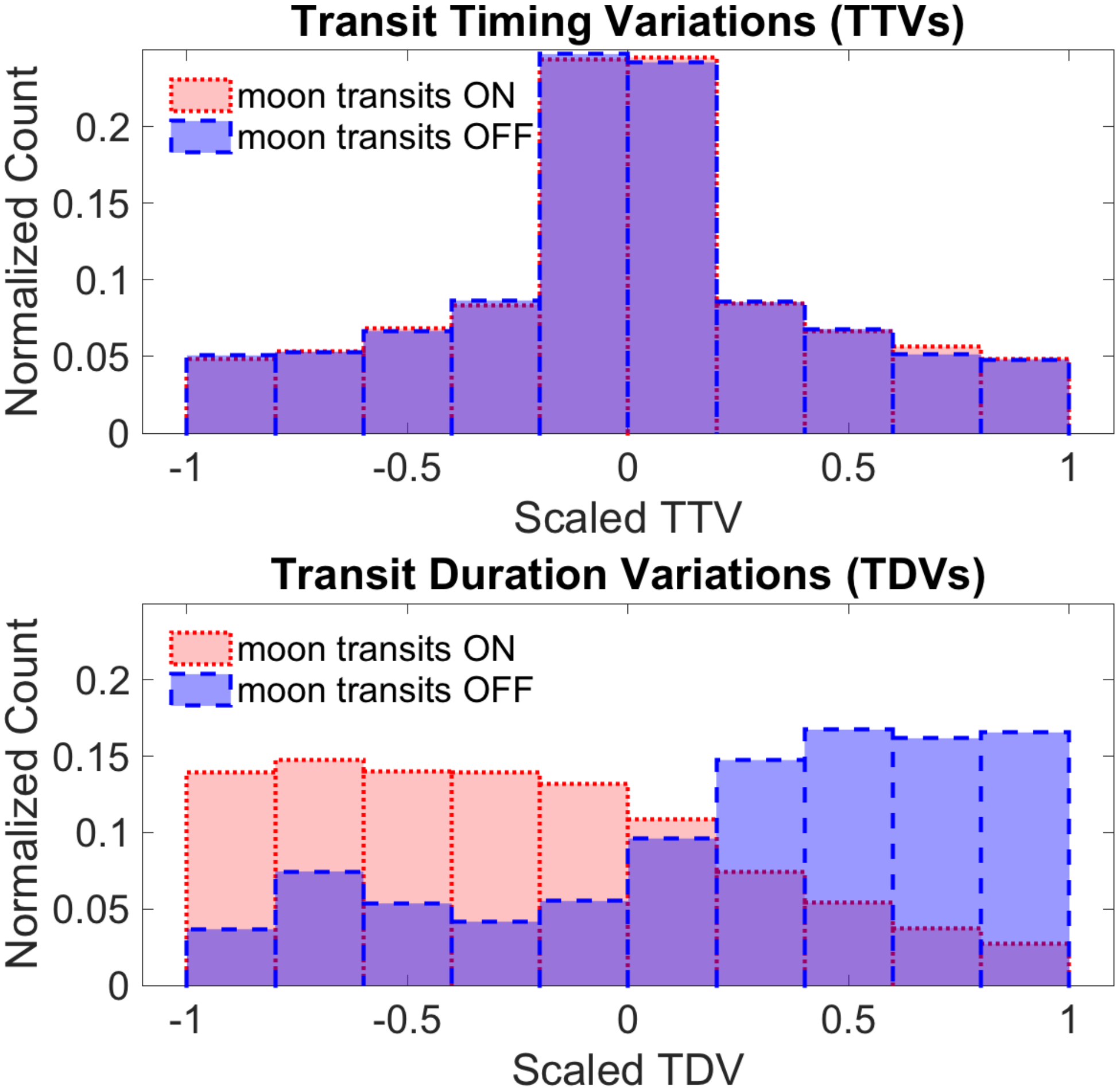}
%\caption{Different type of planet orbits in binaries}
%\caption{\bf Observer's view of a transiting exoplanet (blue), its host star (yellow), and exomoon (red). Moons within the gray region will transit the star. Dotted regions of the moon and planet orbits show where those orbits pass behind the projected orbit of the other body.}
%\label{fig:correlation_main}
\end{minipage}
\begin{minipage}{0.95\textwidth}
\vspace{0.3cm}
Left: normalized histograms of TTVs and TDVs, scaled by the maximum amplitude in each simulation and separated to when the moon does transit (moon transits ON, red) and does not transit (moon transits OFF, blue). The TTVs and TDVs are calculated in the $n$-body simulations presented in Sect.~\ref{subsec:accuracy}, only taking the 626/1000 simulations with at least one missed moon transit. Right: same TDV results but separated into small (top), moderate (middle) and high (bottom) planet impact parameters.
\end{minipage}
\end{minipage}
\begin{minipage}{0.49\textwidth}
\includegraphics[width=0.9\textwidth]{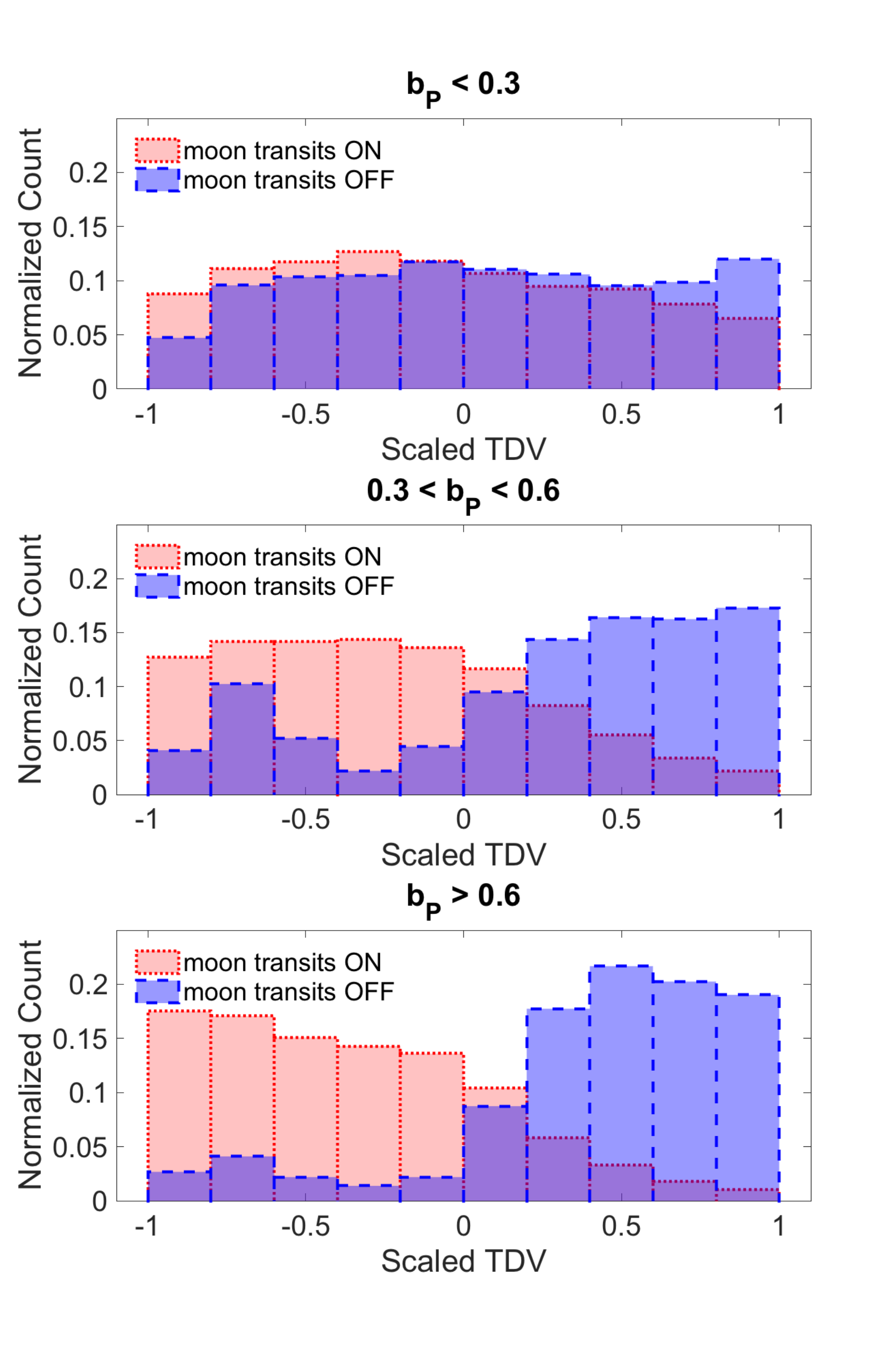}  
%\caption{Different type of planet orbits in binaries}
%\caption{\bf Observer's view of a transiting exoplanet (blue), its host star (yellow), and exomoon (red). Moons within the gray region will transit the star. Dotted regions of the moon and planet orbits show where those orbits pass behind the projected orbit of the other body.}
%\label{fig:correlation_main}
\end{minipage}
\caption{}
\vspace{1cm}
\label{fig:correlation_overall} 
\end{figure*}

\begin{figure*}
\centering {\large (a) $b_{\rm P}=0.1$}
\gridline{\fig{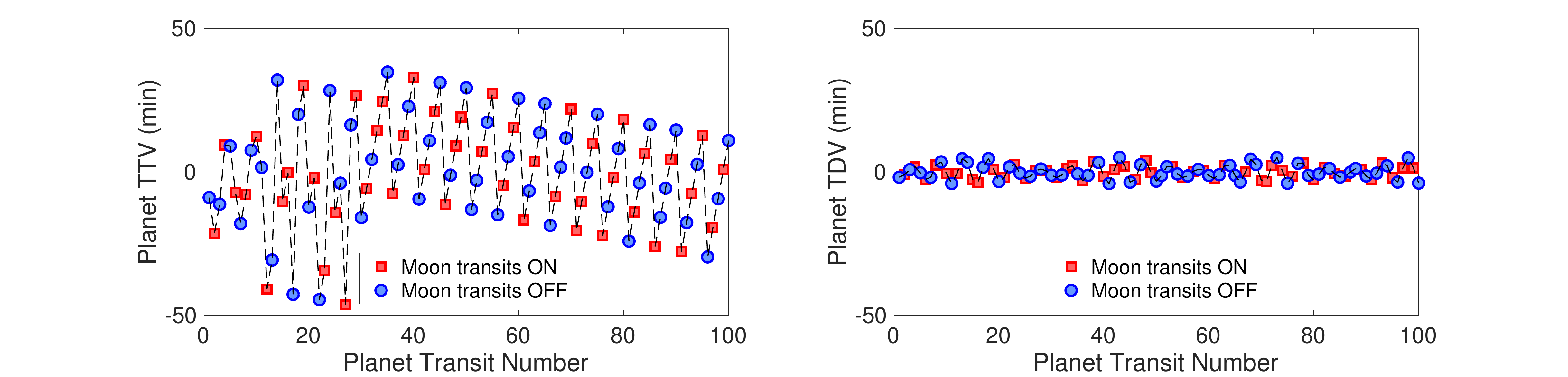}{0.99\textwidth}{}\label{fig:TTV_example_b01}}
\centering {\large (b) $b_{\rm P}=0.4$}
\gridline{\fig{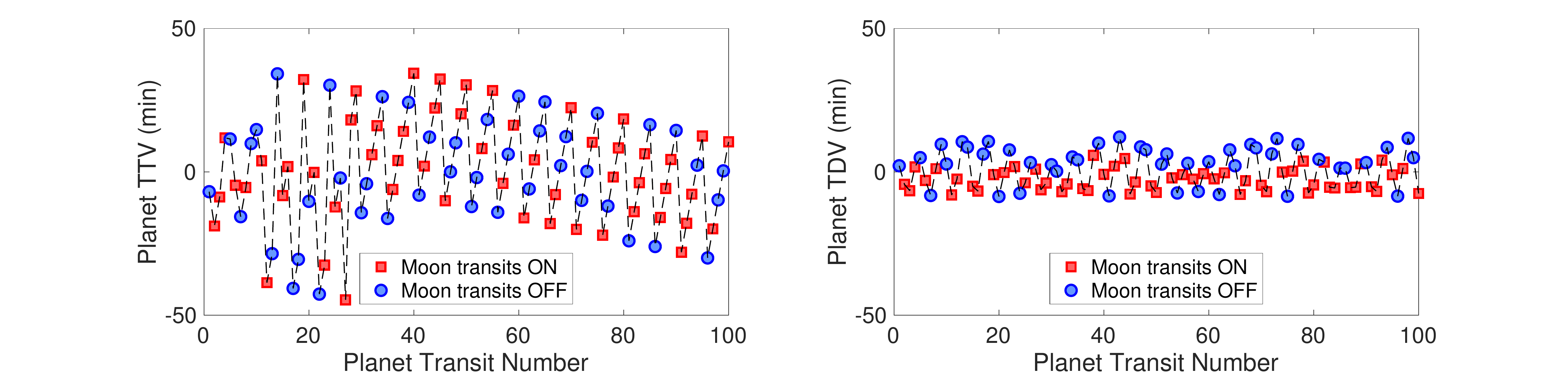}{0.99\textwidth}{}\label{fig:TTV_example_b04}}
\centering {\large (c) $b_{\rm P}=0.7$}
\gridline{\fig{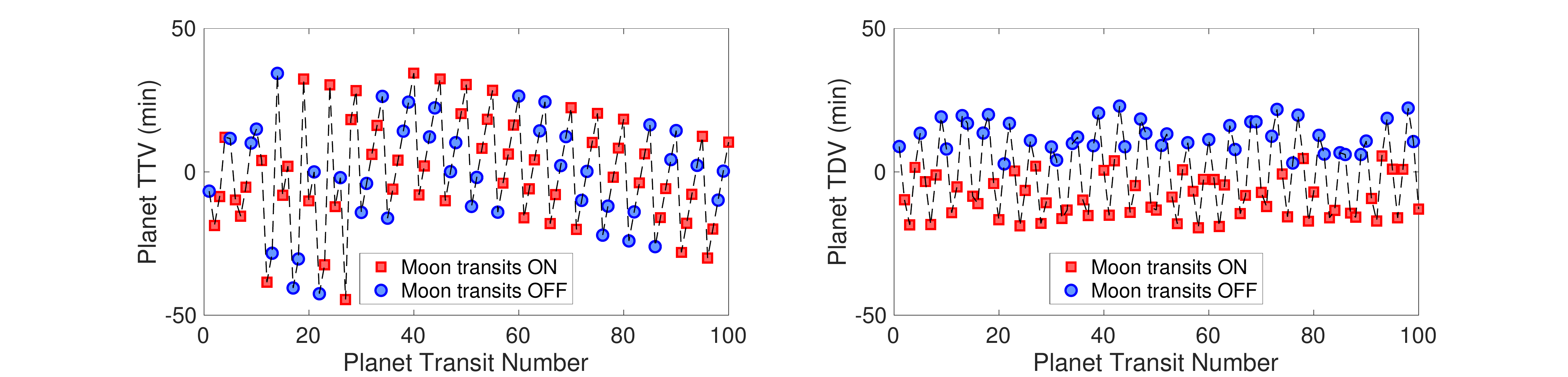}{0.99\textwidth}{}\label{fig:TTV_example_b07}}
\caption{TTVs (left) and TDVs (right) for a $5M_{\oplus}$, $2R_{\oplus}$ moon with $T_{\rm M}=20$ days ($a_M=3.05R_\odot$) around a $1M_{\rm Jup}$, $1R_{\rm Jup}$ planet with $T_{\rm P}=1$ year orbit around a $1M_{\odot}$, $1R_{\odot}$ star, with misalignment of $\Delta I=30^{\circ}$ and planet impact parameters 0.1 (a), 0.4 (b), and 0.7 (c). All simulations start with $\Omega_{\rm M}=0^{\circ}$, $f_{\rm P}=0^{\circ}$, and $f_{\rm M}=50^{\circ}$. Red indicates that moon transits occur, whilst blue indicates that they did not. \label{fig:correlation_examples}}
\end{figure*}

\subsection{Accuracy of the analytic solution}\label{subsec:accuracy}

We run $n$-body simulations for a suite of 1000 randomized transiting planet plus moon systems and calculate a numerical transit probability as the ratio of moon to planet transits. The masses are $m_{\star}=1M_{\odot}$, $m_{\rm P} \in[0.1,3]M_{\rm Jup}$, $m_{\rm M} \in[0.1,17]M_{\oplus}$, using log-uniform distributions. The planet radii are calculated using the mass-radius relation of \citet{Bashi:2017gf}: $R/R_{\oplus}=(m/M_{\oplus})^{0.55}$ for $m/M_{\oplus}<124$ and  $R/R_{\oplus}=(m/M_{\oplus})^{0.01}$ for $m/M_{\oplus}\geqslant124$. The orbital parameters for the planet are $T_{\rm P}\in [200,500]$ days, $e_{\rm P}=0$, $\Omega_{\rm P}=0$, $b_{\rm P}\in [0,0.9]$ and $f_{\rm P}\in [0^{\circ},360^{\circ}]$. The orbital parameters for the moon are $T_{\rm M}\in [1,50]$ days, $e_{\rm M}=0$, $f_{\rm M}\in [0^{\circ},360^{\circ}]$. The mutual inclination is drawn from $\Delta I \in [0^{\circ},40^{\circ}]$. We randomly choose the starting phase of the precession period by calculating $I_{\rm M}$ in Eq.~\ref{eq:moon_incl_over_time} with a uniformly random phase between $\in [0^{\circ},360^{\circ}]$ and $I_{\rm P}$ calculated from the randomly chosen $b_{\rm P}$. We then calculate $\Omega_{\rm M}$ from Eq.~\ref{eq:Delta_I}.

Each simulation is run  over a time span of $100\times T_{\rm P}$ using a fourth-order Runge-Kutta integrator with a fixed step size of 30 minutes, chosen to match {\it Kepler}'s long-cadence observations. Across all 1000 simulations, the median percentage error between the analytic and numerical transit probabilities is 1.2\%. For 626 of the simulations the numerical transit probability is less than 1 (i.e. at least one missed moon transit), and for these simulations the median error is 4.0\%. Contributions to the error include perturbations to the moon's orbit, mean motion resonances, other period-ratio effects which may alias the moon transit sequence, any simplifications in the derivation of Eq.~\ref{eq:moon_transit_probability} and counting statistics of the numerically calculated transit probability. 

\section{Planet transit timing and duration variations}\label{sec:transit_timing}

An isolated, unperturbed planet would transit the star with perfect periodicity, $T_{\rm P}$. However, the presence of the moon can induce TTVs and TDVs on the planet. The main cause is a small ``wobble'' of the planet around the planet-moon barycenter, on top of the planet's larger-scale orbit around the star-planet barycenter. This is a Keplerian effect (i.e., it occurs with static orbits). We briefly diskuss the origin of the barycentric TTVs and TDVs in Sect.~\ref{subsec:timing_origins}, and direct the reader to the seminal papers of \citet{Kipping:2009aa,Kipping:2009bb} for a much more thorough treatment, included detailed analytic equations. A secondary contribution to TTVs and TDVs is from non-Keplerian effects, i.e. perturbations to the orbital elements. We do not diskuss these effects but they are naturally included in our $n$-body simulations. Finally, we do not diskuss the TTVs and TDVs of the moon itself, but they are expected to significantly larger than those of the planet.

\subsection{Origins of barycentric TTVs and TDVs}\label{subsec:timing_origins}

%In this letter we only give brief descriptions for the origin of the TTVs and TDVs and calculate them using $n$-body integrations. For analytic equations, we direct the reader to the seminal papers of \citet{Kipping:2009aa,Kipping:2009bb}. 

%The semi-major axis of the planet's barycentric ``wobble'' is $a_{\rm W} = a_{\rm M}m_{\rm M}/(m_{\rm P}+m_{\rm M})$. 

A planet exhibits a TTV when slightly offset along the {\it horizontal} axis (i.e. parallel with its transit chord). This change adds or subtracts to the time taken to reach the transit midpoint. A horizontal offset is induced by the planet's wobble around the planet-moon barycenter. The TTV is calculated as the time taken for the planet to traverse this offset at its orbital velocity around the star of $v_{\rm P,\star}=2\pi a_{\rm P}/T_{\rm P}$.

A planet exhibits a TDV for two different reasons. First, the planet's motion around the planet-moon barycenter has a velocity $v_{\rm P,M}=2\pi a_{\rm M}m_{\rm M}/[(m_{\rm P}+m_{\rm M})T_{\rm M}]$. The {\it horizontal} component of this velocity may be additive or subtractive to $v_{\rm P,\star}$, and hence when the planet transits it may be moving a little faster or slower than average, causing the transit duration to vary. \citet{Kipping:2009bb} called this the ``V-TDV''.

The second cause of a TDV is a {\it vertical} offset of the planet's position (i.e. perpendicular to its transit chord) due to the barycentric reflex motion induced by the moon. This changes $b_{\rm P}$, hence changing $\tau_{\rm P}$ by Eq.~\ref{eq:tauP}. \citet{Kipping:2009bb} called this the ``TIP-TDV''.

\subsection{Connecting TTVs and TDVs with moon transit occurrence}\label{subsec:correlation}

\begin{figure*}
\centering {\large (a)}
\gridline{\fig{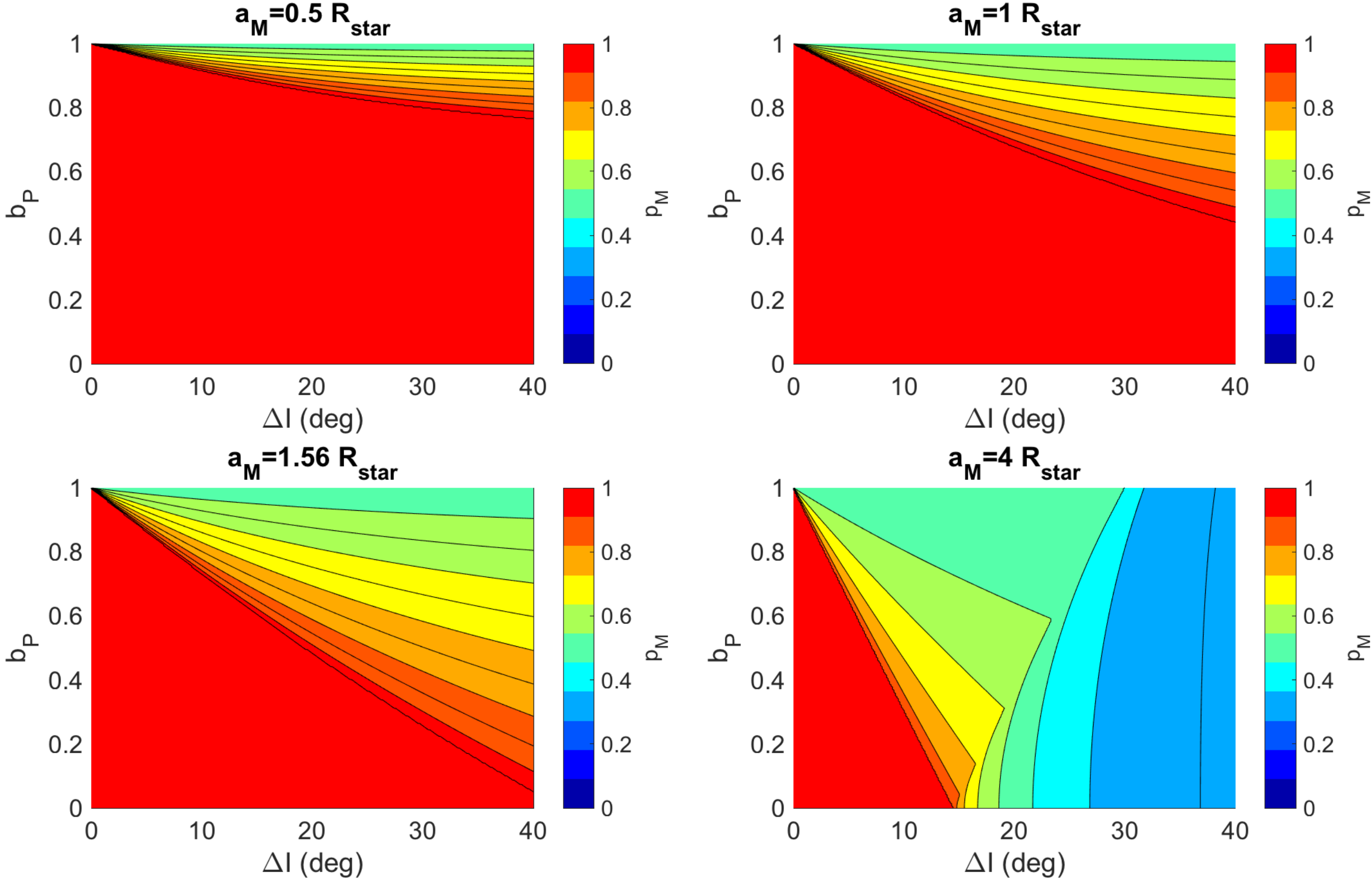}{0.99\textwidth}{}\label{fig:parameter_space_scan}}
\centering {\large (b)}
\gridline{\fig{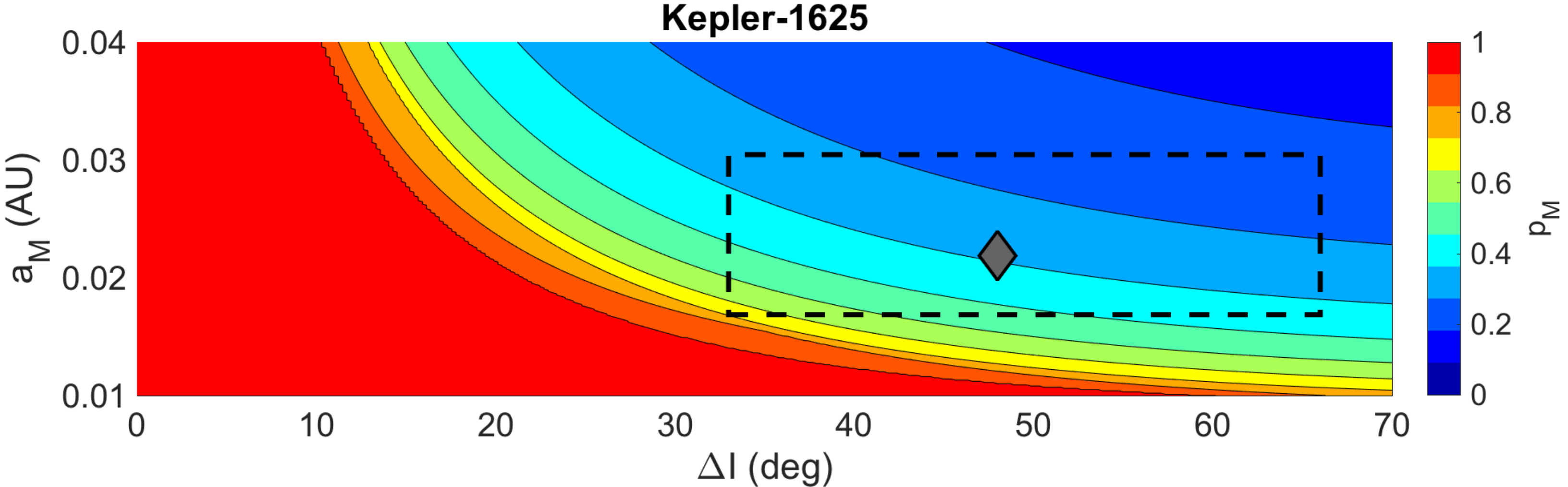}{0.99\textwidth}{}\label{fig:Kepler_1625_application}}
\caption{(a) $p_{\rm M}$ as a function of the planet's impact parameter $b_{\rm P}$, and the mutual inclination, $\Delta I$, for $R_{\star}=1R_{\odot}$ and four different values of $a_{\rm M}$. (b) $p_{\rm M}$ of Kepler-1625b-i using the nominal parameters from \citet{Teachey:2018kj}, where we scan across $\Delta I$ and $a_{\rm M}$. The gray diamond is the best-fitting value and the dashed boxes are $1\sigma$ error bounds. Note that the transit probability is symmetric between prograde and retrograde orbits, and indeed $\Delta I$ could be just as likely $132^{\circ}$ as its noted value here of $48^{\circ}$. Note that in (b) for $\Delta I$ between $40$ and $70^{\circ}$ there will be Kozai-Lidov cycles, which would affect the true $p_{\rm M}$ on the long term but are not accounted for in our equations.}\label{fig:applications}
\end{figure*}

%\begin{figure}
%\gridline{\fig{Correlation_Main.pdf}{0.45\textwidth}{(a)}\label{fig:correlation_main}}
%\gridline{\fig{Correlation_ImpactParameter.pdf}{0.45\textwidth}{(b)}\label{fig:correlation_ImpactParameter}}
%\caption{(a) Normalized histograms of TTVs and TDVs, scaled by the maximum amplitude in each simulation and separated to when the moon does (red) and does not (blue) transit. {\bf The TTVs and TDVs are calculated in the $n$-body simulations presented in Sect.~\ref{subsec:accuracy}, only taking the 626/1000 simulations with at least one missed moon transit.} (b) Same TDV results but separated into small (top), moderate (middle) and high (bottom) planet impact parameters. \label{fig:correlation_overall}}
%\end{figure}

We use the Sect~\ref{subsec:accuracy} $n$-body simulations to test the correlation between moon transits and planet TTVs and TDVs. We only take the 626/1000 simulations which have at least one missed moon transit. For each simulation we calculate numerically the TTVs and TDVs, which we scale by dividing each value by the maximum absolute value for the simulation. We collate the scaled TTVs and TDVs for the simulations, separate them by moon transit occurrence, and show the results in a histogram in Fig.~\ref{fig:correlation_overall} (left).

For  TTVs there is typically no difference between when the moon does and does not transit. There are two main reasons for this. First, occurrence of a moon transit is a function of its vertical position ($y_{\rm M}$), yet the TTV signal is a function of the moon's horizontal position ($x_{\rm M}$). Consider Fig.~\ref{fig:basic_geometry}. A positive $x_{\rm M}$ displaces the planet to the left and hence induces a positive TTV (late transit), and vice-versa. We see that positive $x_{\rm M}$ values correspond to both cases where the moon does and does not transit (only misses above the star). Negative $x_{\rm M}$ values largely correspond to the moon transiting, but there is also a small parameter space for missing transits, both above and below the star. In Fig.~\ref{fig:basic_geometry}, when averaged over all $x_{\rm M}$ there will be preference for missed transits to correspond to positive values of $x_{\rm M}$, and hence positive TTVs. However, this trend will be weak except for small $a_{\rm M}/R_{\star}$, and in that case it would be rare for the moon to avoid transit anyway. The second consideration is that nodal precession of the moon rotates its orbit. After $0.5T_{\rm prec}$ the moon orbit in Fig.~\ref{fig:basic_geometry} will be mirrored horizontally, in which case missed moon transits will now typically correspond to negative values of $x_{\rm M}$. Our $n$-body simulations cover multiple precession periods, and hence any small short-term TTV-moon transit correlations are averaged out.

For TDVs the results contrastingly show a clear difference in the TDV distribution with and without moon transits. This matches Fig.~\ref{fig:basic_geometry}; the moon misses transit when in the uppermost and lowermost parts of its orbit, but the upper region is larger due to the asymmetric vertical offset. When the moon is in this upper region the planet is displaced slightly downward toward the stellar center and hence takes longer to transit (a positive TDV). This does not change throughout the nodal precession period.

The TDV-transit correlation is only prominent when $b_{\rm P}$ is significantly non-zero. In Fig.~\ref{fig:correlation_overall} (right) we split the simulations into $b_{\rm P}\in$ $[0,0.3]$, $[0.3,0.6]$, and $[0.6,0.9]$. The correlation between TDVs and moon transits disappears for small impact parameters. There are two reasons for this. First, for the same vertical offset induced by the moon the change in the path length across the star is less when the planet passes near the stellar center rather than near the limb. Second, at small $b_{\rm P}$ the moon's orbit across the star is nearly symmetric vertically, and hence is nearly equally likely to miss transit above or below the star (unlike in Fig.~\ref{fig:basic_geometry}).

The TDVs for small $b_{\rm P}$ are largely caused by the velocity change effect, which dependent on the horizontal position of the moon and hence is not strongly correlated with the presence of moon transits.

In Fig.~\ref{fig:correlation_examples} we show TTVs and TDVs for three example simulations. The sole change is $b_{\rm P}=0.1$, 0.4 and 0.7. The planet TTV signal remains constant, although the sequence of moon transits changes. The TDV signal at small $b_{\rm P}$ is small in amplitude with no correlation with the moon transits. As $b_{\rm P}$ increases so does the TDV amplitude and the moon transit correlation.

The impact parameter of the planet Kepler-1625b is well constrained to be small: $b_{\rm P}=0.104^{+0.084}_{-0.066}$. We therefore expect TDVs to be small and uncorrelated with missed moon transits, and indeed no TDVs have been observed so far.

\section{Applications}\label{sec:applications}

\subsection{Transit probability of hypothetical exomoon systems}

The transit probability for the moon is a function of $\Delta I$, $b_{\rm P}$, and $a_{\rm M}/R_{\star}$.  Fig.~\ref{fig:applications} (a) shows $p_{\rm M}$ (Eq.~\ref{eq:moon_transit_probability}) over a wide range of parameters: $\Delta I\in[0,40^{\circ}]$, $b_{\rm P}\in[0,1]$, and $a_{\rm M}/R_{\star}=0.5,1,1.56, 4$.

For $a_{\rm M}/R_{\star}<1.56$  the transit probability is 1 except for high values of $\Delta I$ and/or $b_{\rm P}$, where the probability goes to a minimum of 0.5.  The parameter space where $p_{\rm M}<1$ increases as $a_{\rm M}/R_{\star}$ increases. When $a_{\rm M}/R_{\star}>1.56$ the moon's orbit is so wide that its vertical extent exceeds the stellar diameter and $p_{\rm M}<0.5$ for some $\Delta I$ and $b_{\rm P}$.

\subsection{Transit probability of Kepler-1625b-i}

%The candidate Kepler-1625 is proposed to have high $a_{\rm M}\R_{\star}$ and $\Delta I$

%Out of the four determinant factors for the exomoon transit probability - $\Delta I$, $b_{\rm P}$, $a_{\rm M}$ and $R_{\star}$ - for the candidate Kepler-1625 the most poorly characterised are $a_{\rm M}$ and $\Delta I$. 

In Fig.~\ref{fig:applications} (b) we calculate $p_{\rm M}$ for Kepler-1625b-i over a plausible range of $a_{\rm M}$ and $\Delta I$, whilst fixing $b_{\rm P}=0.1$ and $R_{\star}=1R_{\odot}$. Note that when calculating the nominal value of $\Delta I$ we take $\Omega_{\rm M}=0^{\circ}$ and then $\Delta I\approx90^{\circ}-I_{\rm M}$ from Eq.~\ref{eq:Delta_I}. This means $\Delta I=48^{\circ}$, which places the system just within the nominal Kozai-Lidov regime, but the $\Delta I$ and $e_{\rm M}$ variations should be small enough for our equations to remain applicable.

The \citet{Teachey:2018kj} nominal values correspond to $p_{\rm M}=0.4$, although this probability varies significantly within the $1\sigma$ error bounds, and they note that the moon could still have a coplanar orbit, which would mean $p_{\rm M}=1$.

\begin{figure*}  
\begin{center}  
\includegraphics[width=0.99\textwidth]{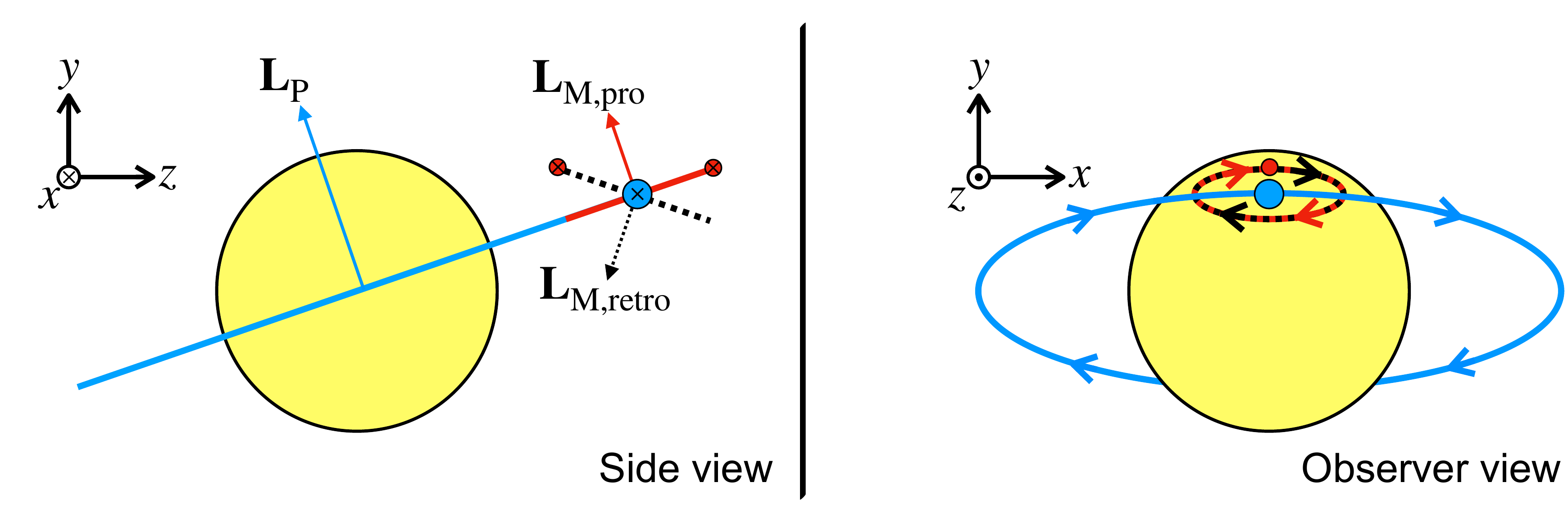}  
%\caption{Different type of planet orbits in binaries}
\caption{An exoplanet orbit (blue) with two different exomoon orbits: the red solid line is prograde and coplanar to the planet, and the black dashed line is retrograde and misaligned to the planet. As seen by the observer (right) there is a degeneracy, as both moons has the same $(x,y)$ position and $(v_x,y_y)$ velocity, despite the side view (left) betraying a clear difference between the two orbits. $\mathbf{L}$ denotes the angular momentum vectors.}
\label{fig:degeneracy}
\end{center}  
\end{figure*} 

\section{Discussion}\label{sec:discussion}

\subsection{Breaking the prograde/retrograde degeneracy}\label{subsec:degeneracy}

%\begin{figure*}  
%\begin{center}  
%\includegraphics[width=0.99\textwidth]{degeneracy.pdf}  
%%\caption{Different type of planet orbits in binaries}
%\caption{An exoplanet orbit (blue) with two different exomoon orbits: the red solid line is prograde and coplanar to the planet, and the black dashed line is retrograde and misaligned to the planet. As seen by the observer (right) there is a degeneracy, as the moon has the same position and velocity, despite the side view (left) demonstrating a clear difference between the two orbits. In the side view arrows show the angular momentum vectors in both possible configurations.}
%\label{fig:degeneracy}
%\end{center}  
%\end{figure*} 

Observations of a moon that orbits a planet on a non-evolving orbit are subject to a degeneracy between prograde ($\Delta I<90^\circ$) and retrograde ($\Delta I>90^\circ$) orbits. This degeneracy is shown in Fig.~\ref{fig:degeneracy}. Two orbits are shown: one in solid red that is prograde and coplanar ($\Delta I = 0^{\circ}$, red solid line), and one in dashed black that is retrograde but misaligned ($90^{\circ} < \Delta I < 180^{\circ}$). Both orbits yield the same projected $x$ and $y$ positions and $v_x$ and $v_y$ velocities of the moon; hence, the Keplerian TTV and TDV phenomenology would be the same. However, the side view (left) reveals a clear difference in the two moon orientations.

This degeneracy may be broken by nodal precession, which would not occur for the coplanar orbit but would for the misaligned orbit. Fortunately, for a moon that orbits at a fair fraction of its planet's Hill sphere, precession will be rapid, revealing the magnitude of the misalignment in just tens of orbits of the planet.  Therefore, the dynamically  evolving character of TDV will betray the prograde or retrograde character of the moon. 

If the planetary impact parameter is low then the ``TIP-TDV'' may be negligible and the magnitude of non-coplanarity may not be enough to break the degeneracy. In this case, higher-order dynamical effects that differ in sign between prograde and retrograde moons may need to be taken into account, as envisioned by \citealt{Lewis:2014rm}. Two alternative methods for breaking the degeneracy, practical only with Extremely Large Telescopes, were diskussed by \cite{Heller:2014as}.

\subsection{The prevalence of large TTVs for long-period gas giants}

According to the transit times of Table S3 of \citet{Teachey:2018kj}, the planet Kepler-1625b has a mean absolute deviation from a constant-period model, normalized by the orbital period --- a ``scatter'' --- of $s_{O-C}/T_{\rm P}=2.40\times10^{-5}$. The timings have a median error bar normalized by the orbital period of $\sigma/T_{\rm P} = 1.55\times10^{-5}$. For the TTV measurements of \citet{Holczer:2016si}, the data are more precise than that for 40 planets with $T_{\rm P}>100$ days. Of those 40, 15 planets have larger TTV scatter, i.e., $s_{O-C}/T_{\rm P}>2.40\times10^{-5}$, and all of these are deemed significant at $\log p<-8.8$. The large amplitude and period of these signals makes them likely due to planet-planet perturbations. We conclude that Kepler-1625b may very likely have a TTV signal due to  additional planets, which may be confused for exomoons, or at least contaminate the exomoon TTV signal. A repeated photometric transit signal of the exomoon, rather than the TTV induced on the planet, is likely a more reliable signature. 

\subsection{Overlapping moon and planet transits}

There are two possible scenarios for overlapping moon and planet transits. First, the moon may be entirely in front of or behind the planet, in which case the photometric signal would be identical to that of an isolated planet transit and the moon would be hidden. Such an event is not explicitly considered in our equations. We estimate it to be rare though, with a likelihood on the order of $\approx R_{\rm P}/a_{\rm M}$ if $I_{\rm M}=I_{\rm P}=90^{\circ}$, and significantly less for inclinations that allow the moon to be offset vertically from the planet at transit. Second, the moon and planet may pass the star at the same time, but with different impact parameters. In this case their photometric dips would be additive and, if telescope precision allowed, a distortion in the transit shape may be detected. Such an event would be covered in our equations for $p_{\rm M}$. Exotic syzygies such as this are treated in more detail in \citet{Kipping:2011tr,Veras:2017rt,Veras:2019he}.

\subsection{Future observing prospects}

The most effective way to confirm and characterize the Kepler-1625 system is through continued transit photometry. Even if the moon only transits $\approx 40\%$ of the time as we predict, additional planet transits will provide new TTV measurements, although probably not new TDV measurements due to the planet's small impact parameter. The next planet transit is scheduled for 2019 May 26. Fig. S18 of \citet{Teachey:2018kj} predicts when the moon will transit. Most of their models show a moon transit before the planet's ingress, but they do not quantify the chance of the moon missing transit \footnote{At the American Astronomical Society Meeting 233, Seattle, 2019 January, Alex Teachey's presentation noted a simulated moon transit chance of 53\% for 2019 May. This is slightly above our 40\% analytic calculation, but our calculations are an average over all transit epochs, not any specific one.}. 

The Transiting Exoplanet Survey Satellite (TESS; \citealt{Ricker:2014qv}) can feasibly observe the planet transit on 2022 July 19 and 2026 June 25, but at Jmag = 14.4 the transit will only be observed at a signal-to-noise ratio of 2.5, which is insufficient for transit timing or moon spotting.

The {\it James Webb Space Telescope} ({\it JWST}) will provide superior photometric precision to  HST \citep{Beichman:2014lh}. From its observing constraints, JWST can observe Kepler-1625 annually from April 22 to November 14, meaning the first planet transits observable with this facility will occur on 2021 October 5, 2022 July 19, and 2023 May 3. With JWST, the transit timing will likely be limited by our abilities to model the granulation features on the stellar surface, which induce significant correlated noise on $\approx 20$ minute timescales given the subgiant nature of this star. Transits of a moon signal of the amplitude and duration claimed by \citet{Teachey:2018kj} will be detectable at the 3$\sigma$ level.

If the moon does not exist, then a binomial test reveals how many non-transits are required to prove this to a certain significance. This assumes that each moon transit would have been detectable and that the transit probability of individual moon transits is independent for each planet transit, which neglects mean motion resonances. The probability of $n$ undetected transits is $p_n = (1-p_{\rm M})^n$. With our estimated $p_{\rm M}=0.4$,  for a $95\%$-confident non-detection we solve $(1-0.95) = (1-0.4)^n$ to obtain $n\sim6$ well-surveyed yet undetected exomoon transits. If the moon does exist, then a similar number of transits would be also be needed to well characterize its orbit.

%If we assume that the detection of individual transits of the moon across the stellar surface are independent at the time of each planet transit, and the probability of detecting an individual moon transit has probability $p_{\rm M}$, then only by continuing to observe multiple non-detections of the system can we rule out the presence of a moon. As each non-detection has probability $1-p_{\rm M}$, the probability of $n$ non-detections is $p_{\rm n} = (1-p_{\rm M})^n$, which can be used to determine the significance of a given number of non-detections. 

\acknowledgments

\noindent 
{\bf Acknowledgements:} We thank the referee for thoroughly reviewing our paper and providing comments which significantly improved its quality.




\bibliographystyle{aa}
\bibliography{../../../library_Kepler1625_arXiv.bib}

%\begin{thebibliography}{}
%
%\bibitem[Astropy Collaboration et al.(2013)]{2013A&A...558A..33A} Astropy Collaboration, Robitaille, T.~P., Tollerud, E.~J., et al.\ 2013, \aap, 558, A33 
%\bibitem[Bertin \& Arnouts(1996)]{1996A&AS..117..393B} Bertin, E., \& Arnouts, S.\ 1996, \aaps, 117, 393 
%\bibitem[Corrales(2015)]{2015ApJ...805...23C} Corrales, L.\ 2015, \apj, 805, 23
%\bibitem[Ferland et al.(2013)]{2013RMxAA..49..137F} Ferland, G.~J., Porter, R.~L., van Hoof, P.~A.~M., et al.\ 2013, \rmxaa, 49, 137
%\bibitem[Hanisch \& Biemesderfer(1989)]{1989BAAS...21..780H} Hanisch, R.~J., \& Biemesderfer, C.~D.\ 1989, \baas, 21, 780 
%\bibitem[Lamport(1994)]{lamport94} Lamport, L. 1994, LaTeX: A Document Preparation System, 2nd Edition (Boston, Addison-Wesley Professional)
%\bibitem[Schwarz et al.(2011)]{2011ApJS..197...31S} Schwarz, G.~J., Ness, J.-U., Osborne, J.~P., et al.\ 2011, \apjs, 197, 31  
%\bibitem[Vogt et al.(2014)]{2014ApJ...793..127V} Vogt, F.~P.~A., Dopita, M.~A., Kewley, L.~J., et al.\ 2014, \apj, 793, 127  
%
%\end{thebibliography}



\end{document}